\documentclass[prl,aps,twocolumn,superscriptaddress]{revtex4-1}
\usepackage{amsfonts}

\usepackage{dcolumn}
\usepackage{bm}
\usepackage{tikz}
\usepackage[colorlinks=true,citecolor=blue,urlcolor=blue]{hyperref}
\usepackage{amsmath,amsfonts,amssymb,times,natbib}
\usepackage[standard]{ntheorem}

\begin{document}

\title{Spin Quantum Heat Engine Quantified by Quantum Steering}
\author
{Wentao Ji$^{1,3\ast}$,
Zihua Chai$^{1,3\ast}$,
Mengqi Wang$^{1,3\ast}$,
Yuhang Guo$^{1,3}$,\\
Xing Rong$^{1,3}$,
Fazhan Shi$^{1,3}$,
Changliang Ren$^{2\dag}$,
Ya Wang$^{1,3\dag}$,
Jiangfeng Du$^{1,3\dag}$
\\
\normalsize{$^{1}$ CAS Key Laboratory of Microscale Magnetic Resonance and School of Physical Sciences, University of Science and Technology of China, Hefei 230026, China.}\\
\normalsize{$^{2}$ Key Laboratory of Low-Dimensional Quantum Structures and Quantum Control of Ministry of Education, Key Laboratory for Matter Microstructure and Function of Hunan Province, Department of Physics and Synergetic Innovation Center for Quantum Effects and Applications, Hunan Normal University, Changsha 410081, China.}\\
\normalsize{$^{3}$ CAS Center for Excellence in Quantum Information and Quantum Physics, University of Science and Technology of China, Hefei 230026, China.}\\
\normalsize{$^{\ast}$ These authors contributed equally to this work.}\\
\normalsize{$^\dag$ Corresponding author. E-mail: renchangliang@hunnu.edu.cn, ywustc@ustc.edu.cn, djf@ustc.edu.cn }
}

\begin{abstract}
Following the rising interest in quantum information science, the extension of a heat engine to the quantum regime by exploring microscopic quantum systems has seen a boom of interest in the last decade. Although quantum coherence in the quantum system of the working medium has been investigated to play a nontrivial role, a complete understanding of the intrinsic quantum advantage of quantum heat engines remains elusive. We experimentally demonstrate that the quantum correlation between the working medium and the thermal bath is critical for the quantum advantage of a quantum Szil\'{a}rd engine, where quantum coherence in the working medium is naturally excluded. By quantifying the non-classical correlation through quantum steering, we reveal that the heat engine is quantum when the demon can truly steer the working medium. The average work obtained by taking different ways of work extraction on the working medium can be used to verify the real quantum Szil\'{a}rd engine.
\end{abstract}
%\pacs{03.65.Ud,%%Entanglement and quantum nonlocality
%03.65.Ta,%%fundation of quantum measurement
%03.67.Hk%%Quantum communication
%}
%================================================================================

\maketitle

Exploring thermodynamics at the quantum level opens up intriguing possibilities, including
testing information theory in the quantum regime \cite{Landauer,Jaynes,Cox,Bennett,Plenio,Maruyama,delRio,Toyabe,Elouard-1},
the development of quantum fluctuation theorems  \cite{Scully,Plastina,Uzdin,Goold,Benenti,Esposito,Campisi,Sgroi,Micadei,Lostaglio},
and the realization of microscopic quantum heat engines (QHE) \cite{Zou,Bouton,VanHorne,Klatzow,Ono,Zanin,Koski,Koski-2,Koski-1,Manzano,Camati,Peterson,deAssis,Cottet}.
In particular, microscopic QHE may operate more efficiently for work extraction than its classical counterpart by exploring the quantum effect.
Over the years, enthusiastic interests have been devoted to implementing QHE by controlling nonequilibrium dynamics in various microscopic systems, such as
atomic systems \cite{Zou,Bouton},
trapped ions \cite{VanHorne},
solid-state spin systems \cite{Klatzow,Ono},
photonic systems \cite{Zanin},
single-electron transistors \cite{Koski,Koski-2,Koski-1,Manzano},
nuclear magnetic resonance \cite{Camati,Peterson,deAssis},
superconducting qubits \cite{Cottet}, among others.
As quantum coherence is an intrinsic property for quantum systems, previous studies have intensively investigated the role of quantum coherence by using single particles or few-level quantum systems as the working medium.
Recently, some researches show this potential high efficiency \cite{Peterson,Manzano-1,Bresque,Ono,Seah},
while the other investigations show that quantum coherence effects are generally detrimental to reaching bounds on the maximum efficiency and power of these efficient thermal engines \cite{Hovhannisyan,Perarnau,Karimi,Brandner,Pekola}.
Hence understanding the advantage of quantumness in QHE qualitatively and quantitatively remains a central issue from both a fundamental and practical perspective of quantum thermodynamics \cite{Beyer}.

In this work, we report an experimental demonstration of a quantum heat engine that can truly exhibit quantum advantage.
By building a quantum Szil\'{a}rd engine, where quantum correlation exists between the working medium and the thermal bath, we have conclusively identified quantum correlation as a source of quantum advantage for QHE, since the reduced state of its working medium is a Gibbs state that naturally excludes the intrinsic coherence feature.
By quantifying the correlation with quantum steering, we clearly show that an optimized steering-type inequality, which is expressed by the average work over different ways of work extraction on the working medium, can distinguish quantum Szil\'{a}rd engines from classical heat engines. 
The more the quantum steering inequality is violated, the more average work the quantum Szil\'{a}rd engine can output than its classical counterpart.

\begin{figure}
	\centering
		%\fbox
	{\includegraphics[width=1.0\columnwidth]{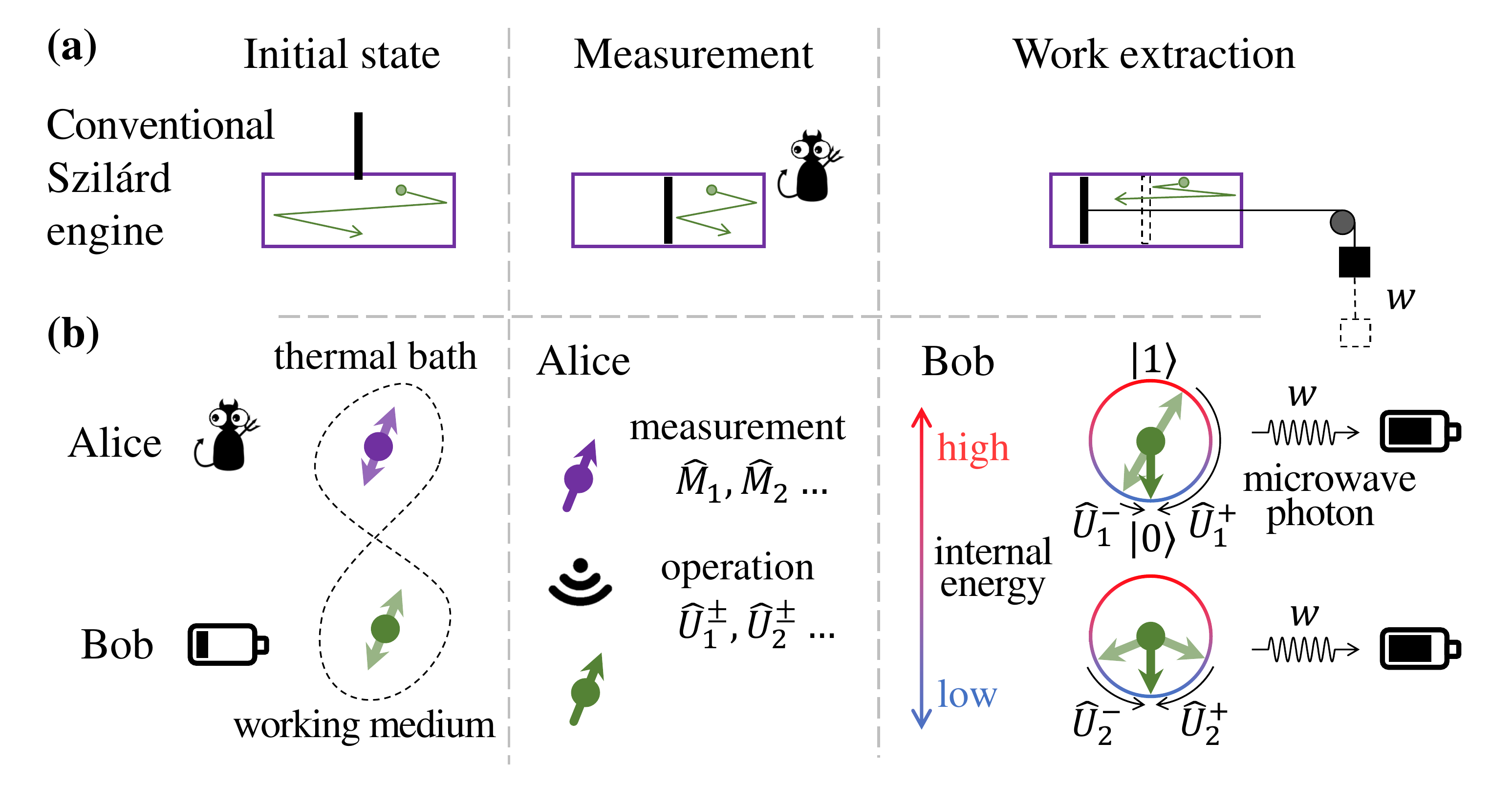}}
	\caption{Conventional and modified Szil\'{a}rd engine.
		(a) For the conventional Szil\'{a}rd engine, the working medium is a single atom that initially stayed in a thermal equilibrium state. A demon measures which half of the box the atom is in. If the atom is in the right half of the box, a movable shutter is put down in the middle. Then the shutter is hung to a load and extracts work from the atom.
		(b) For the modified Szil\'{a}rd engine, both the working medium and the bath are resembled by a single spin qubit respectively. Alice (the demon) prepares the initial state of the whole system, performs measurement $\hat{M_{i}}$ on the bath qubit, and tells Bob the operations $\hat{U}^{\pm}_{i}$ depending on the measurement outcomes $\pm 1$. Bob implements the operations $\hat{U}^{\pm}_{i}$ on the working medium qubit to extract work from its internal energy.
	}\label{fig1}
\end{figure}

%. convensional Szilard engine
The Szil\'{a}rd engine, proposed by Leo Szil\'{a}rd in 1929 \cite{szilard1929,szilard1964}, serves as a prototypical model for understanding the fundamental relation between thermodynamics and information science \cite{Plenio,Maruyama,Toyabe,dunkel,ribezzi,berut}. 
A convensional Szil\'{a}rd engine consists of a single atom as the working medium, which is in the thermal equilibrium within a box~\cite{Maruyama}. 
The demon measures the atom's microstate and controls the single atom doing work, as shown in Fig.~\ref{fig1}(a).
To resemble the conventional Szil\'{a}rd engine, we construct a modified Szil\'{a}rd engine with a quantum system as a proof-of-principal demonstration.
We use one qubit as the working medium, and another qubit as the bath.  
In equilibrium, the working medium is in a Gibbs state~\cite{Maruyama}
\begin{equation}\label{Gibbs}
	\rho_{\mathrm{M}}^{\mathrm{Gibbs}}=\frac{1+\eta}{2}|1 \rangle\langle 1|+\frac{1-\eta}{2}|0 \rangle\langle 0|,
\end{equation}
where $\eta=(e^{-\beta}-1)/(e^{-\beta}+1)$ and $\beta=1/k_{B}T$.
%. quantum Szilard engine, three differences
%The quantum Szil\'{a}rd engine, however, is fundamentally different from its {\color{red} classical counterpart} in three folds.
Due to its quantum nature, the modified Szil\'{a}rd engine is fundamentally different from its conventional counterpart in three folds.
%. measuring the bath 
Firstly, direct quantum measurement of the working medium, in general, destroys its quantum state and disturbs the local thermodynamical situation, like changing the average energy of the working medium \cite{Elouard-1}.
To avoid such disturbance, the demon instead can choose to perform the positive operator-valued measurement on the bath and communicate outcome-dependent energy-conserving operations to the working medium, enabling the subsequent work extraction \cite{Benjamin}.
%. decompositions -> extraction ways
Secondly, for a quantum system of the working medium, a statistical mixture of states allows for infinitely many different ensembles of pure states. Different decompositions may lead to different extraction ways.
%. multiple measurements
Lastly, in general, demonstrating quantum properties requires at least two measurements that do not commute with each other \cite{Beyer,Uola}.
%. summary -> fig.1
After taking these differences into account, Fig.~\ref{fig1}(b) displays the prototypical model of our quantum Szil\'{a}rd engine and the work extraction ways investigated in this work. The demon, named Alice, prepares the initial state of the Szil\'{a}rd engine and decides the energy-conserving operations $\hat{U}^{\pm}_{i}$ depending on the outcomes of the measurement $\hat{M_{i}}$. Bob implements the operations $\hat{U}^{\pm}_{i}$ on the working medium and extracts work from it.

\begin{figure}
	\centering
	%\fbox
	{\includegraphics[width=1.0\columnwidth]{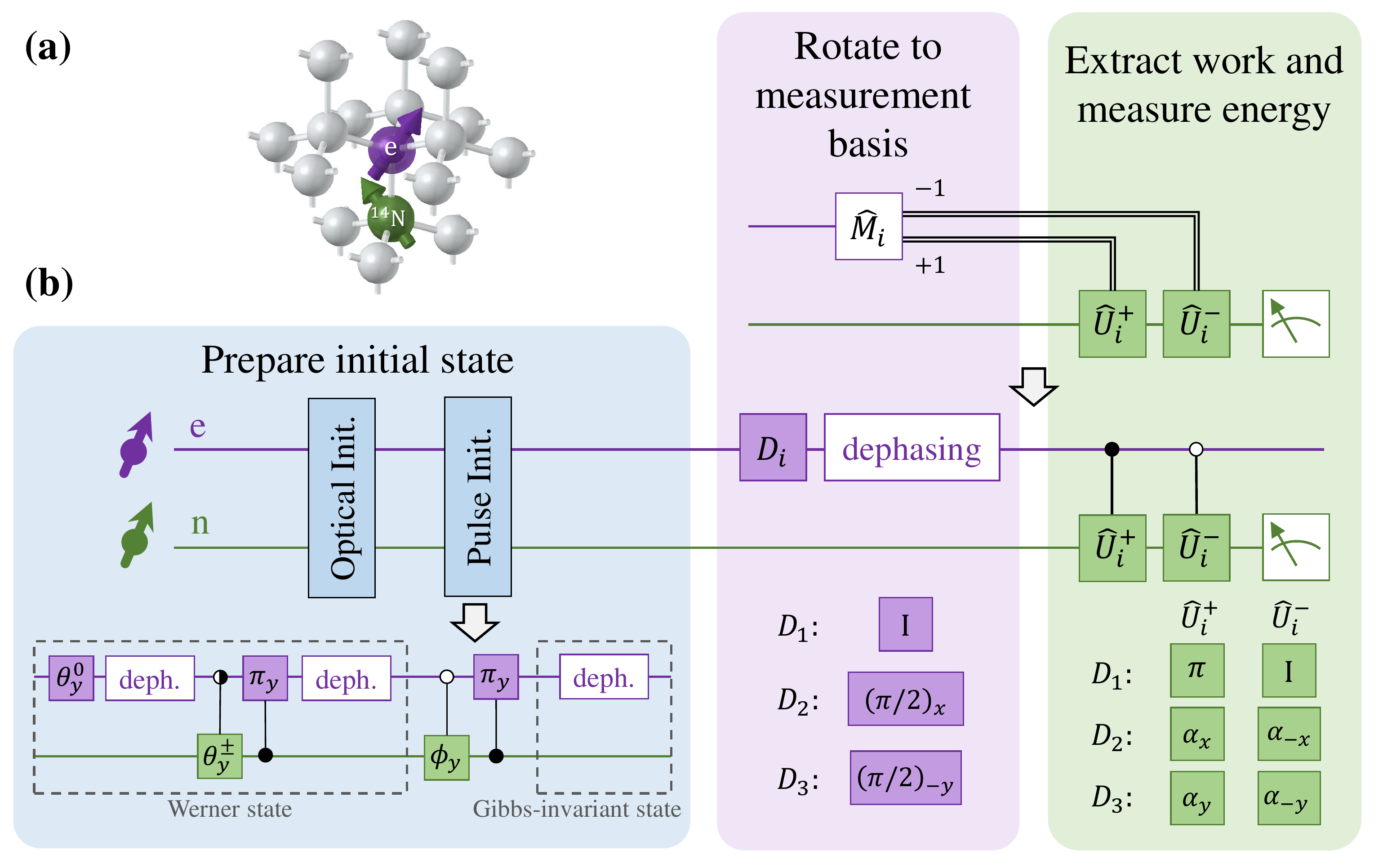}}
	\caption{Experimental circuit for implementing the modified Szil\'{a}rd engine.
		(a) The experimental system: a Nitrogen-Vacancy (NV) center in diamond. The bath and the working medium are resembled by the intrinsic electron spin and Nitrogen nuclear spin of the NV center.
		(b) The quantum circuit for the experimental implementation.
		In the state preparation step, different circuits are used to prepare different types of correlated initial states. The circuit in the left dashed box is for the Werner states, and the right dashed box is for the Gibbs-invariant mixed states. $\theta_y^{0,\pm}$ and $\phi_y$ are rotations along $y$ axis. The rotation angle and the dephasing processes are dependent on $\eta$ and $q$.
		In the measurement step, depending on the decomposition $D_{i=1,2,3}$, the electron spin is measured with $\hat{M}_{1,2,3}=\sigma_{z,y,x}$. This is equivalently realized by the operations labeled by $D_i$ (listed below), which rotates the measurement basis of $\hat{M_{i}}$ to the $\sigma_z$ Pauli basis.
		Finally the work is extracted by the controlled rotations $\hat{U}^{\pm}_{i}$ on the nuclear spin, and the final energy of the nuclear spin is measured~\cite{SM}.
	}\label{fig2}
\end{figure}

%. system
We implement this prototypical quantum Szil\'{a}rd engine with a Nitrogen-Vacancy (NV) center in diamond. The intrinsic nuclear spin of Nitrogen is used as the working medium, and the electronic spin of the NV center is used as the bath.
%. 3 key characteristics
High-fidelity spin manipulation \cite{Dolde,Rong}, on-demand decoherence control \cite{ZK-Li} as well as optical readout of spin state \cite{Neumann,Robledo,Q-Zhang} make this quantum system well-suited for demonstrating microscopic heat engines \cite{Klatzow} and understanding the underlying physics. 
The relevant quantum circuit consists of three steps, as shown in Fig.~\ref{fig2}(b). 
%. initial state
The whole system is optically initialized into $|00\rangle$ state and then prepared to a state with specific quantum correlation by a combination of microwave control of electron spin, radio-frequency control of nuclear spin as well as controllable dephasing processes. 
%. measurement and work extraction
The measurement-outcome-dependent operations are realized by the electron-spin-controlled rotation gates (CROT) on the nuclear spin. The measurements are realized by first rotating the electron spin to the corresponding basis and then dephasing the electron spin before implementing the CROT gates. 
%. readout
The extracted work is obtained by reading out the nuclear spin state and calculate energy difference $W = \mathrm{Tr} (\rho_{\mathrm{n,init}}\hat{H}_{\mathrm{M}}) - \mathrm{Tr} (\rho_{\mathrm{n,final}}\hat{H}_{\mathrm{M}})$, with the internal energy of the working medium $\hat{H}_\mathrm{M} = |1 \rangle \langle 1|$~\cite{SM}.

\begin{figure}
	\centering
	%	\fbox
	{\includegraphics[width=1.0\columnwidth]{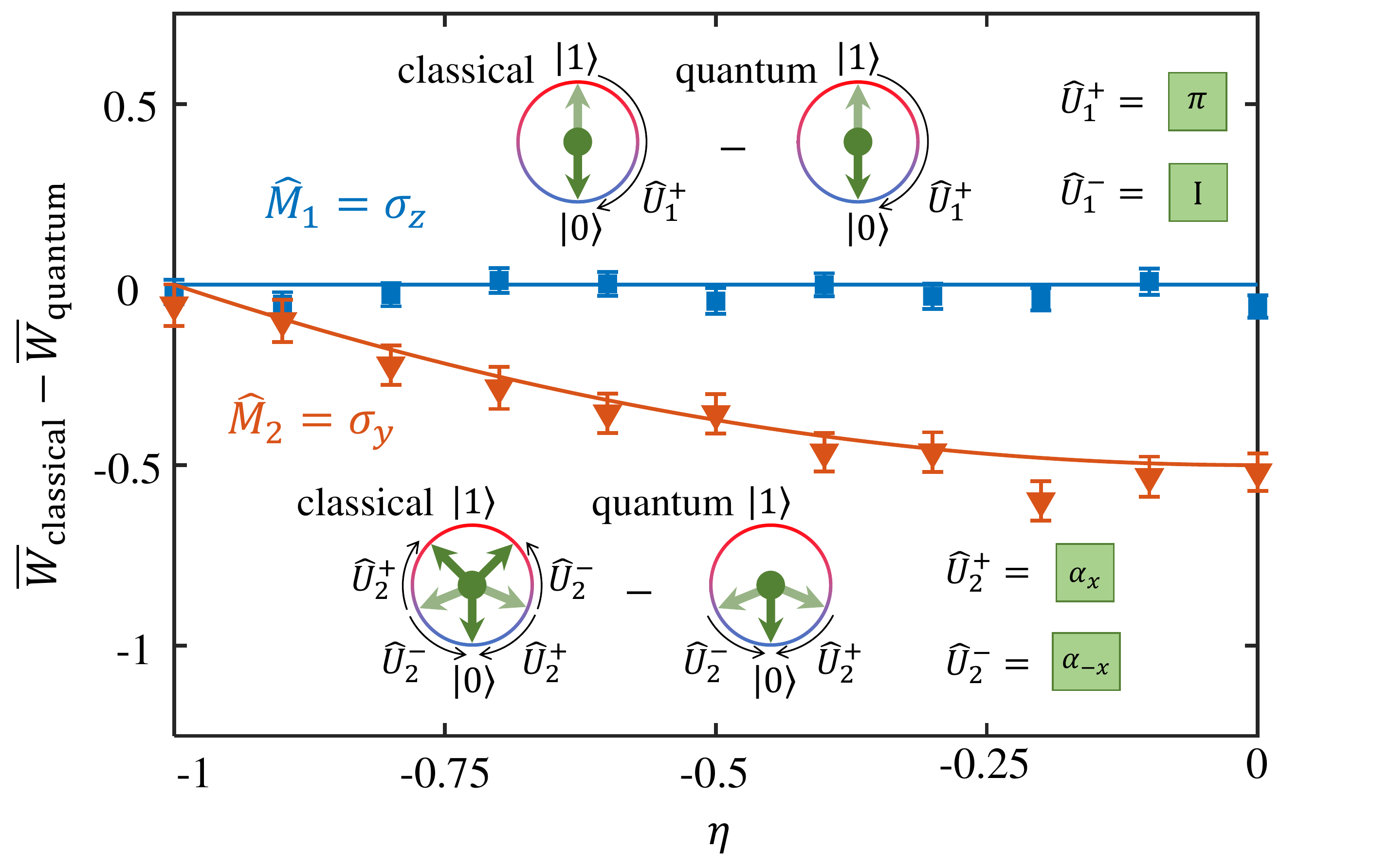}}
	\caption{
		Difference between work extracted from classical and quantum global states. The quantum (classical) global state is a pure entangled state $\rho_1$ (separable state $\rho_2$). The blue (red) data points are the work difference of measurement $\hat{M}_{1}=\sigma_z$ ($\hat{M}_{2}=\sigma_y$), with errorbars representing one standard deviation. The solid lines are the theoretical predictions.
		For the $\hat{M}_{1}$ measurement, work extraction from both the classical and quantum global states are optimal, yielding the same extracted work. While for the $\hat{M}_{2}$ measurement, work extraction from the classical state is no longer optimal, and can extract less work than from the quantum global state.
	}\label{fig3}
\end{figure}

%. understanding
In order to understand the effect of quantum correlation, we first consider two typical global quantum states giving the same thermalized local Gibbs state as illustrated in Eq.~(\ref{Gibbs}). 
%. 2 states
The first state is a pure entangled state of $\rho_{1} = |\psi\rangle \langle\psi|$ with $|\psi\rangle = \sqrt{\frac{1+\eta}{2}} |11\rangle + \sqrt{\frac{1-\eta}{2}} |00\rangle $, and the second state is a separable state of $\rho_{2} = \frac{1+\eta}{2} |11\rangle \langle 11| + \frac{1-\eta}{2} |00\rangle \langle00|$. 
From Bob's side, he can not distinguish these two cases and will expect to extract the same work for the same extraction process. 
%. measurement and operation
In both cases, Alice measures the bath qubit by $\hat{M}_{1} = \sigma_{z}$ or $\hat{M}_{2} = \sigma_{y}$, and chooses from operation $\hat{U}_{1}^{+} / \hat{U}_{1}^{-}$ or $\hat{U}_{2}^{+} / \hat{U}_{2}^{-}$ depending on the measurement result. 
For $\hat{M}_{1}$ measurement, $\hat{U}_{1}^{+}=\sigma_{x}$ for the measurement result $+1$ and $\hat{U}_{1}^{-}=\mathrm{I}$ for the measurement result $-1$. 
For $\hat{M}_{2}$ measurement, $\hat{U}_{2}^{\pm}$ are rotations around the $\pm x$ axis for an angle $\alpha = 2 \arctan \sqrt{ (1+\eta)  / (1-\eta) } $~\cite{SM}. 
%. experimental results
Fig.~\ref{fig3} shows the experimentally extracted work difference between $\rho_2$ (classical) and $\rho_1$ (quantum). 
%. M1, single extraction way
For work extraction with the $\hat{M}_{1}$ measurement, Bob finds no difference between these two Szil\'{a}rd engines constructed by $\rho_{1}$ and $\rho_{2}$. 
The $\hat{U}_{1}^{+}/\hat{U}_{1}^{-}$ operations extract the largest amount of the work by flipping the working medium qubit to the lower energy $|0 \rangle$ state. 
%. M2
For the $\hat{M}_{2}$ measurement, Alice still provides Bob the optimal work extraction for state $\rho_1$, while for $\rho_2$ it is no longer optimal. 
%. averaged
Apparently, the average work of these two measurement-based extractions will unambiguously distinguish the Szil\'{a}rd engines with and without specific quantum correlation.
%. ignore correlation or single extraction
If the investigations only focus on the working medium itself, ignoring the bath-medium correlation, or are limited to a single way of work extraction ($\hat{M}_{1} = \sigma_{z}$ here), which is usually assumed in previous investigations \cite{Campisi}, the effect of quantum correlation will be missing.

%. raise question
One crucial question is whether and when a Szil\'{a}rd engine with quantum correlation can be distinguished from any classical Szil\'{a}rd engines with the same Gibbs state of the working medium.
A better classical strategy may exist for the classical Szil\'{a}rd engine.
However, the optimal average extracted work for such classical Szil\'{a}rd engine is bounded by the local statistical ensemble description, which is a local hidden state model (LHS).
If the average work output from a Szil\'{a}rd engine is greater than the upper bound given by the LHS model, the quantumness of the Szil\'{a}rd engine can be unambiguously identified.
This is analogous to the identification of quantum steering, which is a special kind of quantum correlation \cite{Uola,Jones}.
%. LHS bound, decomposition
Specifically, for the classical Szil\'{a}rd engines with thermalized Gibbs states Eq.~(\ref{Gibbs}) on the working medium, suppose that Bob would like to extract work from three dichotomic pure state decompositions $D_{i=1,2,3}$, of which the bases are given by Bloch vectors $n^{\pm}_{1}=(0,0,\pm1)$, $n^{\pm}_{2}=(0,\mp \sqrt{1-\eta^2},\eta)$ and $n^{\pm}_{3}=(\pm\sqrt{1-\eta^2},0,\eta)$ respectively.
If Bob takes each work extraction $D_{i}$ randomly with the probability $c_{i}$, the locally extracted work is bounded by
\begin{equation}\label{classcial bound}
%\overline{W}_{\mathrm{LHS}}\leq\frac{1}{2}\left(\eta+(c_2+c_3)\eta^2+\sqrt{c^{2}_{1}+(c^{2}_{2}+c^{2}_{3})(1-\eta^2)}\right),
\overline{W} _{\mathrm{LHS}} ^{\mathrm{opt}} = \frac{1}{2} \left(\eta+(c_2+c_3) \eta^2 +\sqrt{c^{2}_{1}+(c^{2}_{2}+c^{2}_{3})(1-\eta^2)} \right),
\end{equation}
namely the local work extraction bound~\cite{SM}.
%. example of rho_1
A quantum strategy for Alice can be to measure $\hat{M}_{1,2,3}=\sigma_{z,y,x}$ according to the decomposition $D_{1,2,3}$ Bob wants to use, and then instruct Bob to perform the corresponding work extraction operation $\hat{U}_{1,2,3}^{\pm}$ depending on the measurement result $\pm1$.
This strategy is optimal for the pure entangled state $\rho_1$, and guarantees the maximal extracted work  $(1+\eta)/2$, exceeding the local work extraction bound. 
%Take the pure entangled state $\rho_1$ for example. No matter which decomposition $D_{1,2,3}$ Bob wants to use, the optimal strategy for Alice will be to measure $\hat{M}_{1,2,3}=\sigma_{z,y,x}$ respectively and instruct Bob to perform the corresponding work extraction operation $\hat{U}_{1,2,3}^{\pm}$ depending on the measurement result $\pm1$.
%This quantum strategy can guarantee the maximal extracted work  $(1+\eta)/2$ for the pure entangled state $\rho_1$, which is greater than its local work extraction bound (see Supplemental Material \cite{SM}). 
%. divide heat engines into quantum and classical
A genuine quantum Szil\'{a}rd engine can therefore be defined by the violation of the work extraction inequality $\overline{W} \leq \overline{W} _{\mathrm{LHS}} ^{\mathrm{opt}}$, which means Bob’s average work output from such engine is larger than what could be obtained from any classical Szil\'{a}rd engine with the same Gibbs state of the working medium. 
It can clearly divide heat engines into two categories, quantum and classical, where no classical statistical description exists for quantum heat engines.

\begin{figure*}
	\centering
	%	\fbox
	{\includegraphics[width=2.0\columnwidth]{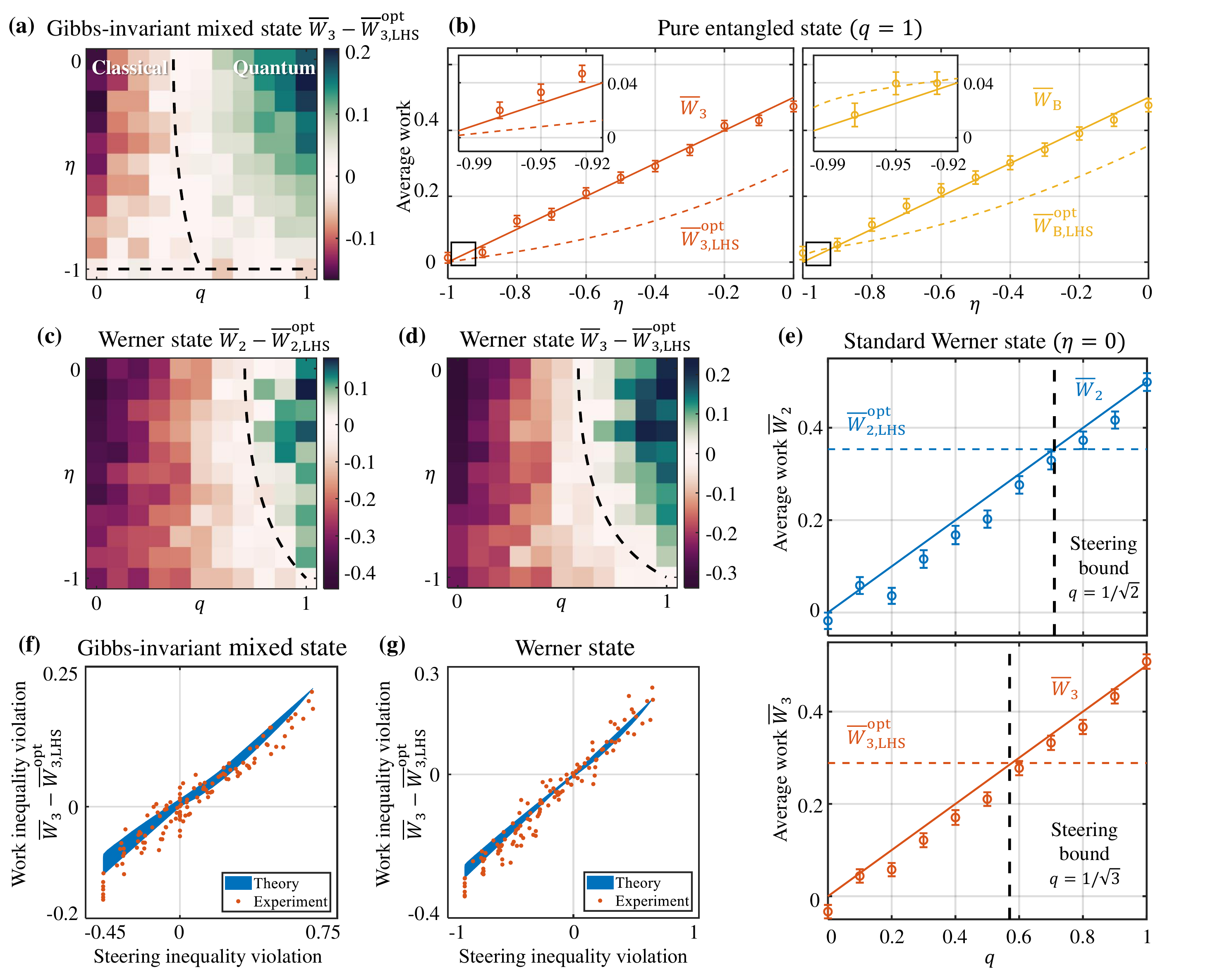}}
	\caption{
		Identification of quantumness with the work extraction inequality.
		(a) Violation of the work extraction inequality $\overline{W}_{3}$ for the Gibbs-invariant mixed states. 
		The color indicates the difference between the extracted work and its local work extraction bound, with the white color indicating the boundary where the work extraction inequality is violated. The dashed black line is the theoretical prediction of the boundary.
		(b) Comparison between the extracted work and its local work extraction bound for the pure entangled states. 
		The data points are experimental results of the extracted work, with the errorbars representing one standard deviation, and the solid lines representing the theortical predictions. The dashed lines are the local work extraction bound. The color of red represents the work extraction with $(c_1,c_2,c_3)=(1/3,1/3,1/3)$, and yellow represents the the work extraction inequality proposed in \cite{Beyer}.
		From the insets, the extracted work of $\overline{W}_3$ violates its local work extraction bounds, whereas $\overline{W}_{\mathrm{B}}$ doesn't, showing that work extraction inequality  $\overline{W}_{3}$ is more effective than $\overline{W}_{\mathrm{B}}$.
		(c,d)  Violation of the work extraction inequality $\overline{W}_{2}$ and $\overline{W}_{2}$ for the Werner states. 
		(e) Comparison between the extracted work and its local work extraction bound for the standard Werner states. 
		For the standard Werner states ($\eta=0$), the work extraction inequalities can be violated when $q>{1}/{\sqrt{2}}$ for $\overline{W}_{2}$ and $q>{1}/{\sqrt{3}}$ for $\overline{W}_{3}$, as marked by the dashed black lines.
		This is consistent with the well-known linear steering inequalities \cite{Saunders}.
		(f,g) Correspondence between quantum steering and work extraction inequality violation \cite{Feng}. 
		The experimental results in (a,d) are plotted as the vertical axis of the data points, and the horizontal axis is the steering inequality violaiton calculated from the corresponding states. 
		The clear positive correlation demonstates the effectiveness of identifying quantum steering with work extraction inequalities.
	}\label{fig4}
\end{figure*}

%. fig4
To demonstrate the classification of quantum and classical Szil\'{a}rd engines, our experiment adopt the Gibbs-invariant mixed states and the Werner states as the global quantum state of the Szil\'{a}rd engine. 
%. define GI state
The Gibbs-invariant mixed states are $\rho_{\mathrm{GI}} = q\rho_{1} + (1-q)\rho_{2}$, where the parameter $q$ tunes the system between the fully quantum case ($\rho_1$ at $q=1$) and the fully classical case ($\rho_2$ at $q=0$). For a given $\eta$, the reduced state of the working medium is always the same Gibbs state, independent of the mixing parameter $q$, as also mentioned in \cite{ Beyer}.
%. fig4(a)
For different $\eta$ and $q$, we measure the average extracted work $\overline{W}_{3}$ ($c_1=1/3,c_2=1/3, c_3=1/3$), and the difference between $\overline{W}_{3}$ and the local work extraction bound $\overline{W} _{\mathrm{LHS}} ^{\mathrm{opt}}$ are shown in Fig.~\ref{fig4}(a). 
It is clearly shown that for some parameters the Szil\'{a}rd engines can output more work than the local work extraction bound $\overline{W} _{\mathrm{LHS}} ^{\mathrm{opt}}$, excluding any classical description, hence the quantum nature is revealed.
The theoretical boundary of the quantum Szil\'{a}rd engines is plotted as the dashed black line.
%. fig4(b)
For the line with $q=1$, $\eta>-1$, the global state is the pure entangled state $\rho_1$. Given that any pure entangled state is steerable \cite{Jones}, we would expect the work extraction inequality to identify the quantumness over the whole line, as shown in the left plot of Fig.~\ref{fig4}(b).
In the right plot of Fig.~\ref{fig4}(b), the results for the work extraction inequality $\overline{W}_{\mathrm{B}}$ as proposed in \cite{Beyer} is also displayed. As shown in the right inset, $\overline{W}_{\mathrm{B}}$ fails to identify quantumness in this region, while in the left inset it is identified successfully, showing the advantage of the inequalities proposed here over $\overline{W}_{\mathrm{B}}$ \cite{Jones,Beyer}. 
%. werner state
The other type of global state is the Werner state, which is the best-known class of mixed entangled states, and first revealed the difference between the notion of entanglement and Bell nonlocality \cite{Werner}. 
%. fig4(c,d)
For Werner states $\rho_{\mathrm{w}} = q\rho_{1} + \frac{1-q}{4} \mathrm{I} \otimes \mathrm{I}$ with different $\eta$ and $q$, the average extracted work $\overline{W}_{2}$ ($c_1=1/2,c_2=1/2, c_3=0$) and $\overline{W}_{3}$ ($c_1=1/3,c_2=1/3, c_3=1/3$) are measured, as shown in Fig.~\ref{fig4}(c,d).
%. fig4(e)
Especially, for standard Werner states ($\eta=0$), the above work extraction inequalities can be violated when $q>{1}/{\sqrt{2}}$ for $\overline{W}_{2}$ and $q>{1}/{\sqrt{3}}$ for $\overline{W}_{3}$, as shown in Fig.~\ref{fig4}(e). These results are consistent with the well-known linear steering inequalities \cite{Saunders}, and strongly demonstrate that the task of clarifying quantum Szil\'{a}rd engine can be one-to-one mapped to the problem of identifying quantum steering when the system is standard Werner states.
%. fig4(f,g)
To further reveal the correspondence between quantum steering and the work extracted from the Szil\'{a}rd engine, we plot the experimental work extraction inequality violation shown in Fig.~\ref{fig4}(a,d) versus the steering inequality violation, which is derived from the all-versus-nothing proof of steering paradox \cite{Feng}. This steering inequality includes naturally the usual linear steering inequality \cite{Saunders} as a special case, and thus can detect more quantum states.
As shown in Fig.~\ref{fig4}(f,g), for both the Gibbs-invariant mixed states and the Werner states, the results show positive correlation, demonstrating that the work extraction inequality is an effective indicator for quantum steering between the working medium and the bath, hence correctly identifies the quantumness of the Szil\'{a}rd engine.
%.
%More detailed discussion and proof for this part are shown in the Supplemental Material \cite{SM}.

\emph{Conclusions-.}
We have experimentally demonstrated a truly quantum Szil\'{a}rd engine in diamond when the demon can ``steer" the working medium where an optimized steering-type inequality that we derived can be violated. Our theoretical and experimental results show that a quantum heat engine which excludes intrinsic coherence feature in working medium, can truly exhibit quantum advantage. We hope our work triggers further studies to generalize our results to the other kind of quantum heat engines.
Our work can be naturally extended to the case of the working medium with higher dimensions.  In the future, it will be interesting to study these heat engines where the working medium is a higher dimensional system. The investigation of quantifying genuine high dimensional quantum steering \cite{Uola,Designolle} can benefit to it. As well, our research can stimulate the bloom of high-dimensional quantum steering.

This work was supported by the National Key R\&D Program of China (Grant No. 2018YFA0306600, 2017YFA0305200), the Chinese Academy of Sciences (Grants No. XDC07000000, No. GJJSTD20200001, No. QYZDY-SSW-SLH004), the National Natural Science Foundation of China (Grant No. 81788101, 12075245, 12104447), China Postdoctoral Science Foundation (Grant No. 2020M671858), Anhui Initiative in Quantum Information Technologies (Grant No. AHY050000), the Natural Science Foundation of Hunan Province (2021JJ10033), the Fundamental Research Funds for the Central Universities, and Xiaoxiang Scholars Programme of Hunan Normal university.

%========================================================================================
%\bibliographystyle{naturemag}%%%%%
%%% From references_heat_engines.bib
%\bibliography{references_heat_engines.bib}
%%%%%

%%% From bbl file generated from references_heat_engines.bib

\end{document}